\documentclass[11pt,twoside]{article} 
\usepackage{fleqn,espcrc1} 
 
\def\be{\begin{equation}} 
\def\ee{\end{equation}} 
\def\bea{\begin{eqnarray}} 
\def\eea{\end{eqnarray}} 

\newcommand{\bee}{\begin{eqnarray}} 
\newcommand{\eee}{\end{eqnarray}} 
 
\title{Light-front model of the kaon electromagnetic   
current
\footnote{ To be published by World Scientific in the proceedings of the
"VIII International Workshop on Hadron Physics,(HADRONS 2002)", 
Bento Goncalves, RS, Brazil, April 14-19, 2002.}} 
 
\author{J. P. B. C. de Melo and Lauro Tomio \\  
Instituto de F\'\i sica Te\'orica, Universidade Estadual Paulista,  
01405-900, S\~ao Paulo, SP, Brazil \\ 
T. Frederico \\  
Departamento de F\'\i sica, ITA, Centro T\'ecnico Aeroespacial,  
12228-900, S\~ao Jos\'e dos Campos, Brazil 
} 
\begin{document} 
\maketitle 
 
The electromagnetic form factor is extracted 
from both components of the electromagnetic current:  
$J^{+}$ 
and $J^{-}$ with a pseudo-scalar coupling of the quarks  
to the kaon. 
In the case of $J^{+}$ there is no pair term  
contribution in the Drell-Yan frame ($q^{+}=0$).  
However, $J^{-}$ component of the electromagnetic  
current  
the pair term contribution is  
different from zero and is necessary include it  
to preserve  
the rotational symmetry of the current.  \\

Light-cone formalisms with model wave-functions have been applied in  
the literature \cite{teren,Pacheco97} to study the pseudoscalar mesons  
\cite{Chung88,Ji90,Frederico92,Cardarelli96,Cardarelli94,Pacheco99,Pacheco2002,Bakker2001}.  
In \cite{Pacheco99}, the form factors in a pseudo-scalar  
coupling model are extracted using the components $J^{+}=J^0 + J^3$ and 
$J^{-}=J^0-J^3$ of the electromagnetic current, in the Drell-Yan frame ($q^+=0$).  
The pair term contribution is present in $J^{-}$, but not in $J^{+}$, where  
the contribution of the pair term vanishes. The pion $J^+$ current, with only 
light-cone valence wave function  results to be equal to the  
covariant calculation, because the pair term is zero.  
In the case of vector particles, even the $J^+$ electromagnetic  
current has contribution from pair terms in order to respect the 
full covariance  \cite{Pacheco991,Pacheco98,Jaus99}.  
In order to satisfy the angular condition for spin one particles, it is 
necessary to consider  pair term in the electromagnetic current $J^+$ 
(see the Ref.\cite{Pacheco991} and references therein).  
Besides the valence contribution to the $J^-$ current, the pair term is 
necessary  for both, pseudoscalar and vector particles to keep the rotational  
symmetry properties of the current in the light-front formalism.  
In order to extract the electromagnetic form factor for the kaon, the 
components $J^{+}$ and $J^{-}$ of the electromagnetic current are used.  
The $J^{(\mu=\pm)}$ components of the electromagnetic current for the 
kaon have contribution, from the quark ($q$) and the  
antiquark ($\bar q$), given by  
\begin{eqnarray} 
J_{q}^\mu (q^2) &=&  \imath e_{q} g^2 N_c \int  
\frac{d^4k}{(2\pi)^4} Tr[S(k-m_{\bar{q}}) 
\gamma^5 S(P^{\prime}-k-m_{q}) \gamma^\mu S(P - k - m_{q})   
\gamma^5 \Lambda(k,P^{\prime}) 
\Lambda(k,P) ] \ ,  \label{j+kaon} \nonumber \\ 
J^\mu_{\bar{q}}(q^2) & = & q \leftrightarrow \bar {q} \ \   
\mbox{in} \ \ J^{\mu}(q^2) \ , 
\end{eqnarray} 
where the number of colors is $N_c=3$, g is the coupling constant  
and $e_q$ ($e_{\bar q}$) is the quark (anti-quark) charge.  
The Breit frame is utilized, where the momentum transfer is 
$q^2=-(\vec q_\perp)^2$, $P^0=P^{\prime \, 0}$ and  
$\vec P^{\prime}_\perp=-\vec P_\perp=\vec \frac{q_\perp}{2}$.  
The function $\Lambda(k,p)=N/((p-k)^2-M^2_{R}+\imath \epsilon)$  
is used in order to regulate the divergent integral, where 
$M_R$ is the regulator mass and $m_{q (\bar q)}$  
is the quark (anti-quark) mass. The function $S(P)$ is the fermion  
propagator.  
The light-front coordinates are defined as  
$k^+=k^0+k^3 \ , k^-=k^0-k^3 \ , k_\perp=(k^1,k^2)\,$.
In the following, the method used in the calculation of the 
pair term, is the one developed in  Ref.{\cite{Pacheco98} 
for a composite boson bound state and in the study of the 
Ward-Takahashi identity in the light-front formalism \cite{Naus98}. 
The contribution of the pair term for $J^{+}$ and $J^{-}$ components  
of the electromagnetic current comes from the matrix elements  
proportional to $k^-$ in both cases (antiquark and quark on-shell).  
 
In order to extract the form factor for the kaon, 
$F_{K^+}(q^2)$, we used both $J^{+}$ and 
$J^{-}$ components of the electromagnetic current.  
 
One can verify that only the on-shell pole 
$\displaystyle \bar{k}^{-}=(f_{1} -\imath \epsilon)/k^+$ 
contribute to the $k^{-}$ integration in the 
interval $0 < k^{+} <  P^{+}$ : 
\begin{eqnarray} 
F^{+}_{\bar q}(q^2) & = &  e_{q} \frac{N^2 g^2 N_c}{P^{+}}  
\int \frac{d^{2} k_{\perp} d k^{+} }{2 \ (2 \pi)^3}  
\frac{ -4 (\bar{k}^{-} k^{+2} - k^{+} k^2_\perp -2  
\bar{k}^- k^+ P^+  
+2 k_\perp P^+ \bar{k}^{-} P^{+}  -k^{+} q/4)}   
{k^+(P^+-k^+)^2 (P^{^{\prime}+}-k^+)^2  
(P^- - \bar{k}^- - \frac{f_2 -\imath \epsilon}{P^+ - k^+}) 
} \nonumber \\ 
& & \frac{-4 k^+ m^{2}_{q} + 8 (k^+ -P^+) m_{q} m_{\bar{q}}  
\ \ \theta(P^{^{\prime}+} -k^+) \theta(k^+ -P^+) } 
{ 
(P^{\prime -} - \bar{k}^{-}  -  
\frac{f_3 -\imath \epsilon }{P^{\prime +} - k^+}) (P^- 
- \bar{k}^{-}  - \frac{f_4 -\imath \epsilon }{P^+ - k^+})  
(P^{\prime -} - \bar{k}^{-} -  
\frac{f_5 -\imath \epsilon }{P^{\prime +} - k^+})}   \ .  \\  
\ F^{+}_{q}(q^2) & = & [ \ q \leftrightarrow \bar{q} \ \  \mbox{in}   
\ F^{+}_{\bar q}(q^2) \ ]   \  ,    
\label{ffactor+} 
\end{eqnarray} 
where  
\ $f_1=k_{\perp}^{2}+m^2_{\bar q}$,   
$f_2=(P - k)_{\perp}^{2}+m^2_{ q}$,   
$f_3=(P^{\prime } - k)_{\perp}^{2}+m^2_{ q}$,   
$f_4=(P-k)_{\perp}^{2}+M^2_{R}$ ,   
$f_5=(P^{\prime}-k)_{\perp}^{2}+M^2_{R}$ \ . 
 
 
In the case the quark is on-shell, the pole contribution is  
$\displaystyle \bar{k}^{-}=(f_{6} -\imath \epsilon)/k^+$  
and $f_6=k_{\perp}^{2}+m^2_{ q}$.

The light-front wave function for the kaon appears 
after the $k^{-}$ integration 
\begin{eqnarray} 
\Phi^i_{\bar q}(x,k_\perp)=\frac{1}{(1-x)^2} \frac{N}{(m^2_{K^+}-M^2_{0}) 
(m^2_{K^+}-M^2_{R})}      
\ ,   
\label{fupi} 
\end{eqnarray}  
where $x=k^+/P^+$. The function $M^2_{R}$ is   
\begin{equation} 
M^2_{R}=M^2(m_{\bar{q}},M_R)=\frac{k^2_\perp+m^2_{\bar q}}{x}+ 
\frac{(P-k)^2_\perp+M^2_{R}}{1-x}-P^2_\perp \ \ ,  
\ \ \ \ ( \bar{q} \leftrightarrow q )  
\ . 
\end{equation} 
The free quark mass square is given by 
$M^2_{0}=M^2(m_{\bar q},m_{\bar q}), \ ( \bar{q} \leftrightarrow q )$.  
For the final wave-functions  
($ \Phi^f_{\bar q}$ and $\Phi^f_{q}$) is necessary exchange  
$P \leftrightarrow P^{\prime  }$.

The expression obtained for the electromagnetic form  
factor in terms of the wave function initial $(\Phi^i_{\bar q})$ and final  
$(\Phi^f_{\bar q})$ is 
\begin{eqnarray} 
F^{+}_{\bar q}(q^2)&=& e_{q} \frac{N^2 g^2 N_c}{ P^+}  
\int \frac{d^{2} k_{\perp} d x}{%
2(2 \pi)^3 x } \bigg[ -4 \Big( f_1 x P^+   
-x P^+ k^{2}_{\perp} - 2 f_1  P^+  +  
2 k^{2}_{\perp} P^{+}  +  \frac{f_1 P^+}{x}  - \frac{x P^+ q^2}{4} \Big)  
\nonumber \\  
&  &  -  \frac{4 f_1 P^{+}}{x} + 8 P^{+} (x - 1) m_{q} m_{\bar{q}} -  
4 x P^+ m_{q}^2 \bigg] 
\theta(x) \theta(1-x) \ \Phi^{*f}_{\bar q}(x,k_{\perp})  
\Phi^i_{\bar q}(x,k_{\perp})  \ , \\  
\ F^{+}_{q}(q^2) & = & [\ q \ \leftrightarrow \ \bar{q}  \  \mbox{in}   
\ F^{+}_{\bar q}(q^2) \ ] \  \ .    
\label{form}  
\end{eqnarray} 
 
The final expression for the  
electromagnetic form factor obtained with $J^{+}$  is the sum of   
two contributions from the quark and the antiquark currents: 
\begin{equation} 
F_{K^+}^{+}(q^2)=F_{q}^{+}(q^2)+F_{\bar{q}}^{+}(q^2) \ ,    
\end{equation} 
where the normalization is given by $F^{+}_{K^+}(0)=1$. 
This expression is free of the pair term contributions, 
due the fact the pair term contributions for the $J^{+}$ component of 
the electromagnetic current is zero  
in the interval (II) \cite{Pacheco99}.  
The calculation of the kaon electromagnetic form factor  
in the light-front with  $J^+$, without pair term, give the same 
result as the covariant one (see Fig.\ref{fig1}).

The contribution to the electromagnetic form factor obtained with $J^{-}$ 
from the interval (I) \ ($0 < k^+ < P^{+}$) is given by 
{\small \begin{eqnarray} 
F^{- (I)}_{ q}(q^2) & = & 2 \imath  e_{q} \frac{N^2 g^2 N_c }{2 P^{+}}  
\int \frac{d^{2} k_{\perp} d 
k^{+} d k^-}{ \ ( 2 \pi)^4}  
\frac{ -4 (k^{-2} k^{+} -  
k^{-} k^{2}_\perp - 2 k^- k^+ P^+  
+2 k^{2}_\perp P^+ +  k^{+} P^{+2}}   
{ k^+ (P^{^{\prime}+}-k^+)^2 (k^- -  
\frac{f_6 - \imath \epsilon }{k^+})} \times \nonumber \\ 
&  & \frac{  - k^{-} q^2/4)  
-4 k^{-2} m^{2}_{\bar q} + 8 (k^-  - P^+) m_{q} m_{\bar{q}} }  
{(P^- - k^- - \frac{f_2 -\imath \epsilon }{P^+ - k^+}) 
(P^{\prime -} - k^- - \frac{f_3 -\imath \epsilon }{P^{\prime +} - k^+}) (P^- 
- k^- - \frac{f_4 -\imath \epsilon }{P^+ - k^+}) (P^{\prime -} - k^- -  
\frac{f_5 -\imath \epsilon }{P^{\prime +} - k^+})} \    
\label{ffactor2-}  
\end{eqnarray} }  
where $f_{2}=(P-k)_{\perp}^2+m^2_{q}$ and  
$\displaystyle 
k^{-}=\frac{k_{\perp}^2+m^2_{q}}{k^+}=f_6/k^+$. 
The second contribution comes from the antiquark current: 
{\small 
\begin{eqnarray} 
F^{-(I)}_{\bar{q}} (q^2) &=& -2 \imath e_{\bar{q}}  
\frac{N g^2 N_c}{2 P^+}  
\int \frac{d^{2} k_{\perp} d 
k^{+} d k^-}{2 \ (2 \pi)^4}  
\frac{ -4 (k^{-2} k^{+} -  
k^{-} k^{2}_\perp - 2 k^- k^+ P^+  
+2 k^{2}_\perp P^+ + k^+ P^{ + 2} }   
{ k^+ (P^{^{\prime}+}-k^+)^2 (k^- - \frac{%
f_1-\imath \epsilon }{k^+})} \times  \nonumber \\ 
&  & \frac{ - k^{-} q^2/4 ) -4 k^{-2} m^{2}_{q} +  
8 (k^+ -P^+) m_{q} m_{\bar{q}} }  
{(P^- - k^- - \frac{f_2 -\imath \epsilon }{P^+ - k^+}) 
(P^{\prime -}-k^- -\frac{f_3 -\imath \epsilon }{P^{\prime +} - k^+})  
(P^- - k^- - \frac{f_4 -\imath \epsilon} 
{P^+ - k^+}) (P^{\prime -} - k^- - \frac{%
f_5 -\imath \epsilon }{P^{\prime +} - k^+})}   
\label{ffacto1-}  
\end{eqnarray} 
} 
where $\displaystyle 
k^{-}=\frac{k_{\perp}+m^2_{\bar{q}}}{k^+}=f_1/k^+$.  
 
After the integration in $k^{-}$  
the  
electromagnetic form factors are    
\begin{eqnarray} 
F^{- (I)}_{\bar q}(q^2) & = &  e_{q} \frac{N^2 g^2 N_c }{P^{+}}  
\int \frac{d^{2} k_{\perp} d x}{ \ 2 (2 \pi)^3 x }  
\bigg[ 
-4 \Big( \frac{f_{1}^{2}}{x P^+} -  
\frac{f_1 k^{2}_\perp}{x P^+} - 2 f_1 P^+  
+2 k^{2}_\perp P^+ +  x  P^{+3} - \frac{f_1 q^2}{ 4 x P^+}  
\Big) \nonumber \\ 
&  &    
- \frac{4 f_1 m^{2}_{q}}{x P^+} + 8 (\frac{f_1}{xP^+} -P^+)  
m_{q} m_{\bar{q}} \bigg]  
\theta(x) \theta(1-x) \ \Phi^{*f}_{\bar q}(x,k_{\perp})  
\Phi_{\bar q}^{i}(x,k_{\perp})  \ . \\  
\ F^{-(I)}_{q}(q^2) & = & [ \ q \  
\leftrightarrow \ \bar{q}  \  \mbox{in}   
\ F^{-(I)}_{\bar q}(q^2) \ ]   \ .    
\label{final2-}  
\end{eqnarray}  
 
When using  $J^-$ 
to extract the electromagnetic form factor,   
besides the contribution of the interval (I), the pair term  
contributes to the  electromagnetic  
form factor in the interval (II) ($P^+ < k^+  < P^{\prime +}$). 
The pair term contribution for the form factor, as shown in  
Refs.\cite{Pacheco99,Pacheco2002}, is given by $F^{-(II)}(q^2)$. 
The final expression for the electromagnetic form factor for the kaon, 
extracted from $J^{-}$ is  
\begin{equation} 
F_{K^+}^{-}(q^2)=\big[F^{-(I)}_{q}(q^2) + F^{-(I)}_{\bar{q}}(q^2) + 
F^{-(II)}(q^2) 
\big]  
\label{final-} \ ,  
\end{equation} 
which is normalized by the charge, $F_{K^+}^{-}(0)=1$.

The parameters of the 
model are the constituent quark masses $m_q=m_{u}=0.220$ GeV, 
$m_{\bar{q}}=m_{\bar{s}}=0.419$ GeV, and 
the regulator mass $M_R=0.946$ GeV, which are adjusted to fit  
the electromagnetic radius of the kaon.  
With these parameters, the calculated electromagnetic 
radius of the kaon is $<r^2_{k^+}>=0.354$ $fm^2$,  
that is very close to the experimental radius $<r^2_{k^+}>=0.340$  
$fm^2$ \cite{Amendolia86}.  
The electromagnetic form factor is presented in Fig.\ref{fig1}.  
Due to the fact that $J^{+}$ does not 
have the light-front pair term contribution, the electromagnetic form  
factor from Eqs.(2) and (3) results equal to the one 
obtained in a covariant calculation.  
In the case of  $J^{-}$, the light-front calculation with Eqs. (11) and (12) 
is quite different from the covariant results, as shown in 
Fig.\ref{fig1}. 
After the inclusion of the pair term in the form factor calculated 
with $J^{-}$, Eq. (13), there is a complete agreement between the 
light-front and the covariant calculations.
  
In conclusion, the $J^{+}$ and $J^{-}$ components of the 
electromagnetic current for the kaon are obtained in the light-front 
and in the covariant formalisms, in a constituent quark model.  
In the case of  $J^{-}$, the inclusion of the pair 
term is essential to obtain the agreement between the covariant 
and the light-front calculations of the kaon electromagnetic form factor.  
 
Our thanks to the Brazilian agencies FAPESP and CNPq for   
financial support.   \\ \\  
 
\vspace{-1.cm}  
 
\begin{figure}[h] 
\vspace{6.5cm} 
\includegraphics{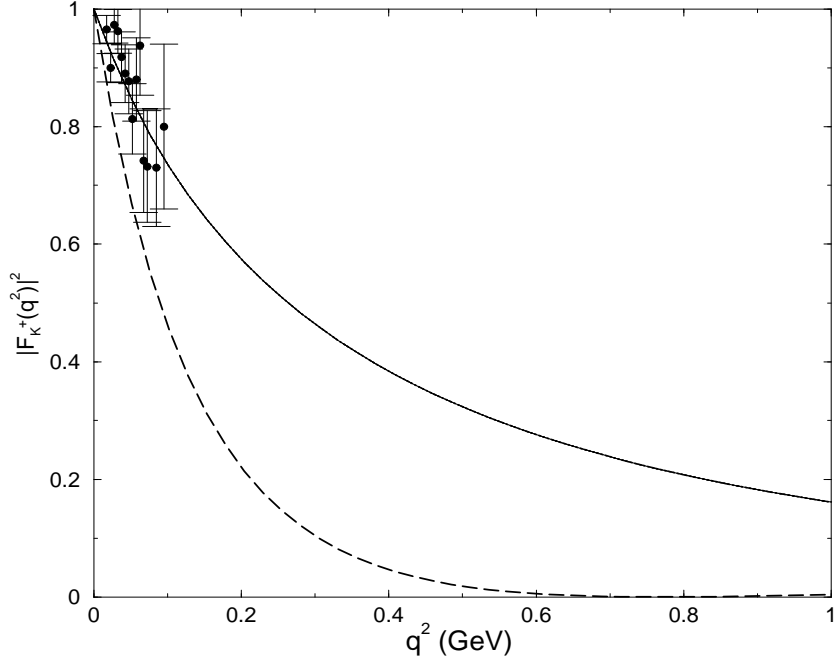} 
\vspace{2.0cm} 
\caption{Kaon ($K^+$) form factor calculated within the covariant  
and the light-front formalisms with $J^{+}$ and $J^{-}$. 
The dashed line give the results from $J^{-}$ without the 
light-front pair term. Adding the pair term to it, we obtain 
the result given by the solid line, which coincides with 
the light-front and covariant calculations with $J^{+}$. The 
experimental data comes from Refs. [15,16].} 
\label{fig1} 
\vspace{1.0cm}  
\end{figure} 
  

\end{document}